\documentstyle[11pt, aaspp4]{article} 

\newcommand{\Teff} {T_{\rm eff}} 

\newcommand{\vsini}{v \sin i}
\newcommand{\kms}{\, {\rm km} \, {\rm s}^{-1}}

\newcommand{\etal}{{\it et al.\ }}
\newcommand{\e}{$\pm \;$}

\newcommand{\unit}[1]{\, {\rm #1}}

\lefthead{Behr et al.}
\righthead{Rotation of M13 BHB stars}

\begin{document}

\title{A New Spin on the Problem of HB Gaps: \\ Stellar Rotation along the Blue Horizontal Branch of Globular Cluster M13\altaffilmark{1,2}}

\author{Bradford B. Behr\altaffilmark{3},
	S. G. Djorgovski\altaffilmark{3},
	Judith G. Cohen\altaffilmark{3}, \\
	James K. McCarthy\altaffilmark{4},
	Patrick C\^ot\'e\altaffilmark{3, 5},
	Giampaolo Piotto\altaffilmark{6},
	Manuela Zoccali\altaffilmark{6}
}

\altaffiltext{1}{Based on observations obtained at the
	W.M. Keck Observatory, which is operated jointly by the California 
	Institute of Technology and the University of California.}
\altaffiltext{2}{Based on observations with the NASA/ESA {\it Hubble Space Telescope}, obtained at the Space
	Telescope Science Institute, which is operated by AURA, Inc., under NASA contract NAS 5-26555.}
\altaffiltext{3}{Palomar Observatory, Mail Stop 105-24,
	California Institute of Technology, Pasadena, CA, 91125}
\altaffiltext{4}{PixelVision, Inc., 4952 Warner Avenue, Suite 300, Huntington Beach, CA, 92649}
\altaffiltext{5}{Sherman M. Fairchild Fellow}
\altaffiltext{6}{Dipartimento di Astronomia, Universit\`a di Padova, I-35122 Padova, Italy}

\begin{abstract}

We have determined the projected rotational velocities of thirteen blue horizontal-branch (BHB) stars in the globular
cluster M13 via rotational broadening of metal absorption lines. Our sample spans the photometric gap observed in the
horizontal branch distribution at $\Teff \simeq 11000 \unit{K}$, and reveals a pronounced difference in stellar rotation
on either side of this feature---bluewards of the gap, all the stars show modest rotations, $\vsini < 10 \kms$, while to
the red side of the gap, we confirm the more rapidly rotating population ($\vsini \simeq 40 \kms$) previously observed
by Peterson \etal (1995). Taken together with these prior results, our measurements indicate that a star's rotation is
indeed related to its location along the HB, although the mechanism behind this correlation remains unknown.  We explore
possible connections between stellar rotation and mass loss mechanisms which influence the photometric morphology of
globular cluster HBs.

\end{abstract}



\section{Introduction}

A number of unresolved issues in post-main-sequence stellar evolution revolve around the nature of stars on the
horizontal branch (HB). The HB stars in globular clusters are readily identified by their position in a cluster's
color-magnitude diagram, and are assumed to have the same age and initial composition as the other stars in the cluster.
The photometric properties of the HB are very sensitive to stellar composition and structure, so detailed study of the
HBs of many different clusters allows us to test models of post-main-sequence stellar evolution.

The distribution of stars along the horizontal branch differs from cluster to cluster. This variety in the HB color
morphology is primarily a function of cluster metallicity (Sandage \& Wallerstein 1960; Sandage \& Wildey
1967), with metal-rich clusters tending to have short red HBs, and metal-poor clusters exhibiting predominantly blue HBs. Some other
parameter (or set of parameters) is evidently also at work, however, as clusters with nearly identical metallicities can
show very different HB color distributions (van den Bergh 1967; Sandage \& Wildey 1967). This `second parameter' was
initially thought to be cluster age (Searle \& Zinn 1978; Lee, Demarque, \& Zinn 1994; Stetson, Vandenberg, \& Bolte
1996), but several alternative or additional second-parameter candidates have since been suggested, including helium
abundance and mixing (Sweigart 1997), CNO abundance (Rood \& Seitzer 1981), central concentration of the cluster (Fusi
Pecci \etal 1993, Buonanno \etal 1997), and distribution of stellar rotation rates (Peterson \etal 1995). Most of these
proposed mechanisms influence the HB morphology by regulating either the core mass, or the amount of mass loss that a
star experiences on the upper red giant branch (RGB). Greater mass loss results in a smaller hydrogen envelope
surrounding the $\simeq 0.5 M_\odot$ helium core, and the star appears on the zero-age horizontal branch (ZAHB) with a
higher photospheric temperature and higher surface gravity. There has been some success in synthesizing the observed HB
distributions by evolving stellar models up the RGB and on to the HB with a mean mass loss $\Delta M \simeq 0.15
M_\odot$ and a dispersion in mass loss $\sigma \simeq 0.01-0.03 M_\odot$ (Lee, Demarque, \& Zinn 1990), but this
approach fails to reproduce details of the complex color morphology seen in many clusters.

Particularly intriguing are the gaps which appear in the blue HB sequence of many globular clusters,
including  M15 (Buonanno \etal 1983), M13 and M80 (Ferraro \etal 1997, Ferraro \etal 1998), NGC 2808 (Sosin \etal 1997), NGC 6229 (Catelan
\etal 1998), and NGC 6273 (Piotto \etal 1999), as well as the metal-poor halo (Newell 1973). High-precision photometry reveals narrow regions of the HB color axis
which are underpopulated at a statistically significant level, and which seem to occur at similar locations in different
clusters, strengthening the claim that they are real features, and not just statistical artifacts (Ferraro \etal 1997,
Piotto \etal 1999). The presence of such gaps poses a major challenge to HB theory, as it is difficult to envision a
single mass-loss mechanism which implements such narrow `forbidden zones' in the envelope mass distribution.
Alternatively, the gaps may mark the boundaries between separate, discrete populations of HB stars, which differ in
their origin or evolution. We then might expect to see some additional qualitative difference
between stars on either side of the gap.

In order to evaluate how various stellar characteristics change as a function of position along the HB and from cluster
to cluster, we have undertaken a program to measure rotation rates and chemical abundances of blue horizontal-branch
(BHB) stars in M3, M13, M15, M92, and M68 via high-resolution echelle spectroscopy. M13 (NGC 6205) is one of the closest
and best-studied globulars, with $(m-M) = 14.35$ mag (Peterson 1993) and a metallicity [Fe/H]~$= -1.51$~dex measured
from red giant abundances (Kraft \etal 1992). Its BHB extends from the blue edge of the RR Lyrae gap to the
helium-burning sequence (a `long blue tail'), interrupted by one or more gaps (Ferraro \etal 1997). We have observed
thirteen of M13's BHB stars in the range $7000 \unit{K} < \Teff < 20000 \unit{K}$, specifically in order to span the
most obvious gap, which is located at $\Teff \simeq 11000 \unit{K}$ and labelled `G1' by Ferraro \etal In this paper, we
describe the measurement of projected stellar rotation ($\vsini$) for this sample; the abundance anomalies also
exhibited by these stars have been reported previously (Behr \etal 1999).

Our measurements build upon prior work by Peterson, Rood, and Crocker (1995, henceforth P95), who undertook a
spectroscopic survey of BHB stars in M13, using the O I triplet at 7775~\AA\ to evaluate [O/Fe] and $\vsini$. In the 29
BHB stars for which $\vsini$ was determined, they found rotations as fast as $40 \kms$, although the small number of
these higher $\vsini$ values strongly suggests that the true distribution of rotational $v$ is bimodal, with roughly a
third of M13's BHB stars being `fast rotators' ($\simeq 40 \kms$) and the other two-thirds being `slow rotators'
($\simeq 15 \kms$). The existence of such fast rotation is difficult to explain. The progenitors of these HB stars were
solar type or later, and are expected to have shed most of their angular momentum via magnetically-coupled winds early
in their main-sequence lifetimes, reaching the $v < 2 \kms$ observed in the Sun and other similar Population I dwarfs. 
Assuming solid-body rotation and a homogenous distribution of angular momentum per unit mass, Cohen \& McCarthy
(1997) estimate that the HB stars should be rotating no faster than $10 \kms$, but this prediction clearly disagrees
with P95, as well as prior measurements in other clusters (Peterson 1983, 1985a, 1985b) and Cohen \& McCarthy's own
$\vsini$ findings in globular cluster M92. Pinsonneault, Deliyannis, \& Demarque (1991) explain the anomalously fast rotation via core-envelope
decoupling---the stellar core retains much of its original angular momentum, while only the envelope is braked, and the rapidly rotating core
is then revealed after much of the envelope is lost on the RGB. Such core behavior would explain the fast
rotators, although the model does not explain why only a third of the HB stars would show such an effect, nor does it
agree with the slow core rotations inferred for young stars (Bouvier 1997, Queloz \etal 1998) and the solar interior
(Corbard \etal 1997, Charbonneau \etal 1998).

The existence of this quickly-rotating population is suggestive when M13 is compared to other clusters. The other
two globular clusters in the P95 study, M3 and NGC 288, were found to have significantly lower maximum rotation among
their BHB stars, $\vsini < 20 \kms$ in nearly all cases. The HB distributions of these two clusters are also somewhat
less blue than M13---neither show any HB population at $\Teff$ above 14500 K---so the obvious inference is that the fast
stellar rotation is somehow related to the presence of a blue tail. Mengel \& Gross (1976) point out a possible
mechanism by which this might occur---fast core rotation in pre-HB stars nearing the tip of the red giant branch would
forestall the helium flash, such that more envelope mass is lost to stellar winds, and the star ends up further to the
blue when it reaches the HB. This mechanism would function on a star-by-star basis, so one would thus expect to see
higher rotation in the hotter stars, as the quickly-rotating core is revealed by progressively greater amounts of RGB
mass loss. No such variation in $\vsini$ as a function of $\Teff$ was seen in the P95 results, although the rapid
disappearance of the oxygen lines above 10000 K limited the temperature coverage of their measurements. Since our sample
extends to $\Teff \simeq 19000 \unit{K}$, we are able to test this hypothesis more thoroughly.


\section{Observations and Reduction}

Our spectra were collected using the HIRES spectrograph (Vogt \etal 1994) on the Keck I telescope during four observing
runs on 1998 June 27, 1998 August 20--21, 1998 August 26--27, and 1999 March 09--11. A 0.86-arcsec slit width yielded a
nominal $R = 45000$ ($v = 6.7 \kms$) per 3-pixel resolution element. Spectral coverage ran from $3940-5440$~\AA\ ($m =
90-66$) for the June observations, and $3890-6280$~\AA\ ($m = 91-57$) for the August and March observations, with slight
gaps above 5130~\AA\ where the free spectral range of the orders overfilled the detector. We limited frame exposure
times to 1200 seconds, to minimize susceptibility to cosmic ray accumulation, and then coadded three frames per star.
$S/N$ ratios were on the order of $50-90$ per resolution element, permitting us to accurately measure even weak lines in
the spectra.

Nine of the thirteen stars in our sample were selected from photometry of archival HST WFPC-2 images of the center of
M13, specifically reduced for this project. The program stars were selected to be as isolated as possible; the HST
images showed no apparent neighbors within $\simeq 5$ arcseconds. The seeing during the HIRES observations was
sufficiently good ($0.8-1.0$ arcsec) to avoid any risk of spectral contamination from other stars. The other four M13 HB
program stars were taken from the sample of P95, to permit a comparison to published values. All of the stars from P95
are located in the cluster outskirts, where crowding is not so problematic. Positions, finding charts, photometry, and
observational details for the entire target list will be provided in a later paper (Behr 1999b).

We used a suite of routines developed by J.~K. McCarthy (1988) for the FIGARO data analysis package (Shortridge 1993) to
reduce the HIRES echellograms to 1-dimensional spectra. Frames were bias-subtracted, flat-fielded against exposures of
the internal quartz incandescent lamp (thereby removing much of the blaze profile from each order), cleaned of cosmic
ray hits, and coadded. Thorium-argon arc spectra provided wavelength calibration, and arc line profiles were used to
determine the instrumental broadening profile (see below). Sky background was negligible, and 1-D spectra were extracted
via simple pixel summation. A 10th-order polynomial fit to line-free continuum regions completed the normalization of
the spectrum to unity.


\section{Analysis and Results}

The resulting stellar spectra show ten to over two hundred metal absorption lines each. Even to the eye, the
hotter stars in our sample are clearly slow rotators, with exceedingly sharp, narrow lines. Measurement of $\vsini$
broadening traditionally entails cross-correlation of the target spectrum with a rotational reference star of similar
spectral type, but this approach assumes that the template star is truly at $\vsini = 0$, which is rare. Furthermore,
given the abundance peculiarities that many of these stars exhibit (Behr \etal 1999), it is difficult to find an appropriate spectral
analog. Since we are able to resolve the line profiles of our stars, we instead chose to measure $\vsini$ by fitting the
profiles directly, taking into account other non-rotational broadening mechanisms.

First, an instrumental broadening profile was constructed from bright but unsaturated arc lines for each night of
observation. Those line profiles which deviated most from the mean were successively discarded, until the rms deviation
reached 1\% of peak intensity or less. This technique proved very effective at eliminating blended arc lines, residual
cosmic ray hits, and otherwise non-representative line profiles from the composite profile. The variation in
instrumental broadening over the detector area proved negligible, as the 100 to 200 lines which made up each composite
profile were distributed evenly across the chip. The resulting instrumental profile was almost perfectly Gaussian, with
a FWHM of $6.3 \kms$, slightly narrower than that expected for $R = 45000$. We estimate a thermal Doppler broadening of
$3 \kms$ FWHM, and a microturbulent broadening $\xi = 2 \kms$ from the previous abundance analysis of these stars, so we
convolved the instrumental profile with a Gaussian of $3.6 \kms$ FWHM to account for these effects. This profile was
convolved with a hemicircular rotation profile for the specified $\vsini$ to create the final theoretical line profile.
No attempt was made to include limb darkening, as this should have only minimal impact on the final result.

Each observed line in a spectrum was fit to the theoretical profile using an iterative least-squares algorithm. Four
free parameters were used for fitting: $\vsini$, line center $\lambda_{\rm ctr}$, line depth, and continuum level. Data
and best-fit profiles for three representative lines in each of two stars are depicted in Figure~1. The values for
$\vsini$ from different spectral lines generally agree quite well, as shown in Figure 2.
We removed helium lines, blended lines such as the Mg II 4481 triplet, and extreme outliers (those
more than $3\sigma$ from the mean) from each line list, and calculated the mean and rms error in the mean in standard
statistical fashion.

Rotation results for the 13 stars observed in this study are given in Table 1. To quantify a star's position on
the HB, we use $\Teff$, which is derived from photographic $BV$ photometry for the P95 stars, and from $UV$
photometry for the nine HST-selected stars, as detailed in Behr \etal 1999. (The photographic photometry
temperatures will be refined by CCD photometry of the cluster in the near future, but are adequate for our present
purpose.) In Figure 3, we plot both our results and the results from P95 as a function of temperature, along with
photometric data illustrating the temperature distribution of the HB star population. The $V$ vs. $\Teff$ color-magnitude
diagram in the middle panel, and the accompanying histogram in the lower panel, are based on our reduction of
the archival HST images.


\section{Discussion}

For the four stars in common, our $\vsini$ agree quite well with those of P95. The existence of the
quickly-rotating BHB population, then, seems secure. However, our data clearly show that the fast rotators are
constrained to the cooler end of the BHB, as {\it all} of the stars between 11000 K and 19000 K appear at $\vsini < 10
\kms$. This abrupt change in the rotation distribution at 11000 K (the location of the gap) was hinted at in the prior
measurements, but appears much more clearly as the observations are extended to hotter stars.

The measurement of $\vsini$ by line broadening is inherently statistical, since a small value of $\vsini$ in a single
star might be due to its rotation axis lying close to the line of sight. Given an isotropic distribution in axis
orientation, however, a large $\sin i$ is much more likely than a small one---the probability that $\sin i < 0.25$ is
about 3\%, for example---and with several stars in the sample, the likelihood that the small observed $\vsini$ are due
to chance polar orientation becomes exceedingly slim. We are confident, then, that the hot stars in our sample are truly
rotating no faster than $10 \kms$, and that we are seeing a real difference in stellar rotation characteristics between
the cooler and hotter stars on the BHB.

The nature of this difference, however, seems to argue {\it against} the hypothesis proposed by P95. If the blue end of
M13's HB is populated by stars which have undergone extreme RGB mass loss due to fast core rotation, then that angular
momentum should become evident at the surface and appear as higher $\vsini$. Instead, the hotter stars, which have
suffered greater mass loss, are observed to have slower rotational velocities. The mechanisms that regulate mass loss
are still poorly understood, so it is certainly possible that faster rotation might somehow limit the amount of mass
lost, and keep those stars at the red end of the BHB. A more likely possibility, also mentioned in P95, is that
enhanced mass loss, due to some other second parameter, is very efficient at removing angular momentum from the
convective envelopes of the RGB stars. Those stars which lose more mass, ending up further towards the blue end of the
BHB, would thus be rotating more slowly. In this case, the range of rotation rates would be better described as a
result, not a cause, of the second-parameter phenomenon. Such a mechanism would explain why the hotter stars in our
sample are all slow rotators, but does not account for the presence of the slow rotating population at lower BHB
temperatures.

An alternative hypothesis is that the slow rotators are the `normal' HB stars, and that the rapid rotators redward of
the gap represent the progeny of merged stars, perhaps a subpopulation of blue straggers (BS), similar to the suggestion
by Fusi Pecci \etal (1992) that BS progeny populate the red HB. Such stellar merger products would likely retain excess
angular momentum even through the HB stage, although the issue is complicated by other possible effects of mergers,
including mass loss. A more comprehensive assessment of such anomalously high HB rotation in M92 (Cohen \& McCarthy
1997) and other clusters, and its correlation with blue straggler populations, should be undertaken.

Of even greater significance is the abruptness of the change in rotational signature, and its location along the HB
axis, which closely coincides with the location of the gap G1, as illustrated in Figure 3. Below 11000 K,
the quickly-rotating stars account for roughly a third of the HB population, but they abruptly disappear at $\Teff >
11000 \unit{K}$. In addition, the slowly-rotating segment of the cooler population does appear to have a higher peak
$\vsini \simeq 14 \kms$ than the hotter stars with $\vsini < 10 \kms$. This discontinuous change in the distribution of
$\vsini$ that we observe on either side of the gap supports the hypothesis that the gap separates two
different populations of HB stars, and may provide an important hint towards eventual identification of the mechanism or
mechanisms which differentiate the two populations.

The low $\vsini$ exhibited by the hotter stars is also relevant to the issue of the anomalous photospheric abundances
reported by Behr \etal (1999)---helium underabundance by as much as a factor of 300, and enhancement of metals to
solar and supersolar levels, probably due to diffusion mechanisms in the radiative atmospheres of the hotter,
higher-gravity stars. Slow stellar rotation, and smaller resulting meridional circulation currents, are likely to be
prerequisites for the diffusion to be effective (Michaud \etal 1983). If the observed HB color morphologies are due to
changes in atmospheric structure brought on by these metal enhancements, as suggested by Caloi (1999) and Grundahl \etal
(1999), then the change in maximum $\vsini$ along the HB could prove closely linked to the appearance of the gaps.
Further theoretical consideration of the influence of stellar rotation on the onset of the diffusion effects and the
impact of diffusion enhancements on photometric colors will be necessary to evaluate this possibility.

Although we are still far from explaining the underlying cause of this gap, or the origin of the anomalously high
rotation rates of the cooler BHB stars, these observations will help to constrain theoretical models, and perhaps
stimulate new ones. Stellar rotation rate clearly has some relation to the second parameter problem, although the nature
of this relation is yet to be determined. Further observations, both in M13 and in other clusters which show similar
gaps, will show whether this rotational discontinuity is a ubiquitous feature of globular cluster BHBs, and how it
is correlated with their morphology.


\clearpage
\acknowledgments

These observations would not have been feasible without the HIRES spectrograph and the Keck I telescope. We are indebted
to Jerry Nelson, Gerry Smith, Steve Vogt, and many others for making such marvelous machines, to the W. M. Keck
Foundation for making it happen, and to a bevy of Keck observing assistants for making them work. SGD was supported, in
part, by the Bressler Foundation, and STScI grant GO-7470. PC gratefully acknowledges support provided by the Sherman M.
Fairchild Foundation. This research has made use of the SIMBAD database, operated at CDS, Strasbourg, France.



\clearpage
 
\begin{figure}
	\epsscale{1}
	\plotone{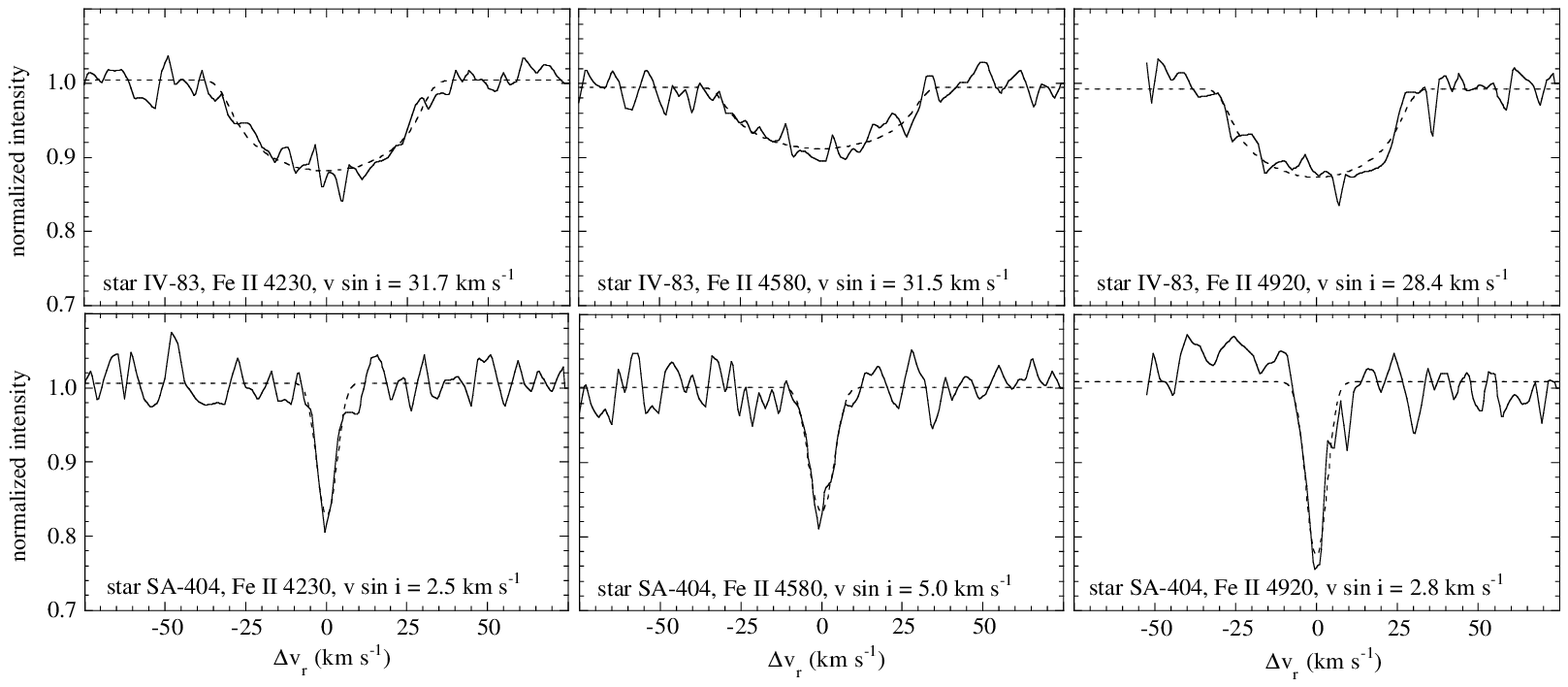}
	\caption{Rotationally-broadened synthetic profiles are fit to observed absorption lines via
		an iterative least-squares algorithm. The top panels show star IV-83, for which we found a mean $\vsini = 32.2 \kms$,
		while the bottom panels depict star SA404, with $\vsini = 3.7 \kms$.}
\end{figure} 

\begin{figure}
	\epsscale{0.5}
	\plotone{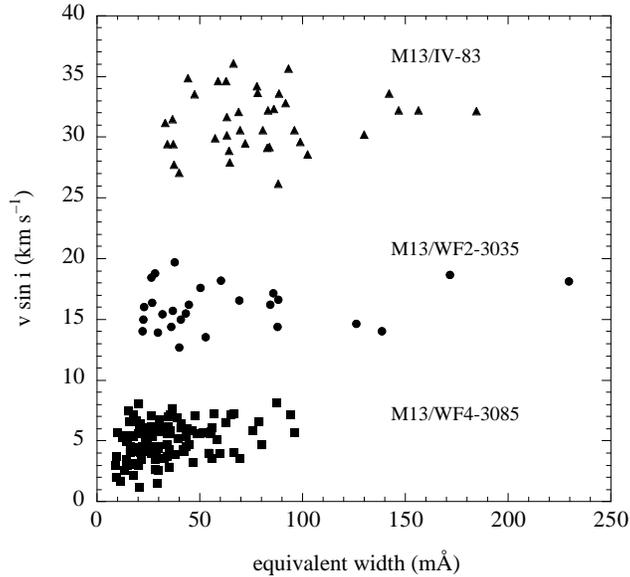}
	\figcaption[fig2.eps]{Derived $\vsini$ for metal absorption lines in each of three stars, plotted as a function of equivalent 
		width $W_\lambda$. The agreement among lines is quite good, and shows no trend in $W_\lambda$. The star WF4--3085 
		exhibits weaker lines than the others because of its higher $\Teff$.}
\end{figure} 

\begin{figure}
	\epsscale{0.5}
	\plotone{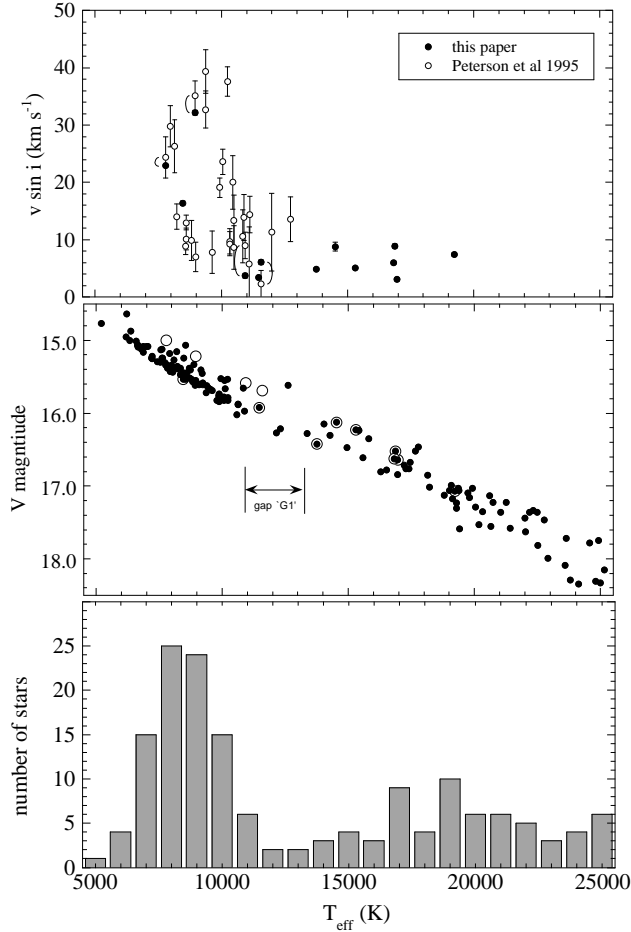}
	\figcaption[fig3.eps]{Projected rotational velocities and color distribution of stars along the HB of M13. In the top
		panel, we plot $\vsini$ for the 13 stars in our sample (solid symbols) and the 29 stars from Peterson \etal 1995 (open
		symbols) as a function of $\Teff$. The curved lines connect measurements of the four stars that are common to both
		samples. In the middle panel is a $V-\Teff$ CMD showing the `gap' at $\Teff \simeq 11000 K$. (Note that the horizontal axis
		is reversed from the normal CMD orientation.) Open circles indicate stars observed as part of our program. In the bottom
		panel is a histogram of star counts, with bins of size 1000~K.}
\end{figure} 


\clearpage 

\begin{deluxetable}{lrcrr}
\tablenum{1}
\tablewidth{0pt}
\scriptsize
\tablecaption{Projected rotational velocities ($\vsini$) for program stars}
\label{tab1}
\tablehead{
				&				&number		&$\vsini$ 			&Peterson \etal 1995		\nl
Star				&$\Teff$ (K)		&of lines		&(km s${}^{-1}$)		&$\vsini$ (km s${}^{-1}$)
}
\startdata
J11				&7780			&67			&22.9	\e 0.2		&24.4	\e 3.6			\nl
WF2--3035		&8470			&26			&16.4	\e 0.4		\nl
IV-83			&8960			&38			&32.2	\e 0.4		&35.1	\e 2.6			\nl
SA404			&10940			&7			&3.7		\e 0.4		&9.0		\e 2.3			\nl
WF3--1718		&11480			&167		&3.4		\e 0.2		\nl
SA113			&11580			&8			&6.1		\e 0.4		&2.3		\e 2.3			\nl
WF4--3485		&13760			&50			&4.9		\e 0.3		\nl
WF2--820			&14520			&46			&8.8		\e 0.7		\nl
WF4--3085		&15300			&102		&5.0		\e 0.1		\nl
WF3--548			&16820			&23			&6.0		\e 0.3		\nl
WF2--2692		&16860			&94			&8.9		\e 0.4		\nl
WF2--2541		&16970			&45			&3.1		\e 0.3		\nl
WF2--3123		&19220			&18			&7.4		\e 0.3		\nl
\enddata
\end{deluxetable}


\end{document}